# A SECURED CRYPTOGRAPHIC HASHING ALGORITHM

Prof. Rakesh Mohanty[#1], Niharjyoti Sarangi[#2], Sukant Kumar Bishi[#3]

[#]*Department of Computer Science and Applications*
*VSSUT, Burla, Orissa, India*
[1]`rakesh.iitmphd@gmail.com`
[2]`nihar.uce@gmail.com`
[3]`sukant.bishi@gmail.com`

*Abstract*— Cryptographic hash functions for calculating the message digest of a message has been in practical use as an effective measure to maintain message integrity since a few decades. This message digest is unique, irreversible and avoids all types of collisions for any given input string. The message digest calculated from this algorithm is propagated in the communication medium along with the original message from the sender side and on the receiver side integrity of the message can be verified by recalculating the message digest of the received message and comparing the two digest values. In this paper we have designed and developed a new algorithm for calculating the message digest of any message and implemented it using a high level programming language. An experimental analysis and comparison with the existing MD5 hashing algorithm, which is predominantly being used as a cryptographic hashing tool, shows this algorithm to provide more randomness and greater strength from intrusion attacks. In this algorithm the plaintext message string is converted into binary string and fragmented into blocks of 128 bits after being padded with user defined padding bits. Then using a pseudo random number generator a key is generated for each block and operated with the respective block by a bitwise operator. This process is iterated for the whole message and finally a fixed length message digest is obtained.

## I. INTRODUCTION

Hash functions were introduced in cryptography in the late seventies as a tool to protect the authenticity of information. Soon it became clear that they were a very useful building block to solve other security problems in telecommunication and computer networks. A cryptographic hash function is a deterministic procedure that takes an arbitrary block of data and returns a fixed-size bit string, the (cryptographic) hash value, such that an accidental or intentional change to the data will change the hash value. The data to be encoded is often called the "message", and the hash value is sometimes called the message digest or simply digests.

A message digest is a code which is created algorithmically from a file and represents that file uniquely. If the file changes, then the message digest changes. In addition to allowing us to determine if a file has changed, message digests can also help to identify duplicate files. The Message Digest is intended for digital signature applications, where a large file must be "compressed" in a secure manner before being encrypted with a private (secret) key under a public-key cryptosystem such as RSA. Other applications of message digest include E-mail security, cyclic redundancy Checksum for files on a network etc.

*Problem Definition:*

The message can be a string of any variable length containing alphabets (both lower case and upper case), digits and special symbols (containing line feeds, form feeds and escape characters). In other words we can write a general formula for the message as:

$$Message = \left((a\text{-}z) + (A\text{-}Z) + (0\text{-}9) + (!\text{ - };)\right)^{*}$$

The message digest should be a string of fixed length containing all the alphabets and special symbols.

$$Message\ Digest = \left((a\text{-}z) + (A\text{-}Z) + (0\text{-}9) + (!\text{ - };)\right)^{n},$$
*Where n = a fixed natural number*

*Motivation:*

While working on MySQL databases and PHP for web applications we came across MD5. We also spotted an application called "Cain n Able" which can crack messages from their MD5 hash value.

- When analytic work indicated that MD5's predecessor MD4 was likely to be insecure, MD5 was designed in 1991 to be a secure replacement.
- In 1993, Den Boer and Bosselaers gave an early, although limited, result of finding a "pseudo-collision" of the MD5 compression function; that is, two different initialization vectors which produce an identical digest.
- In 1996, Dobbertin announced a collision of the compression function of MD5 (Dobbertin, 1996). While this was not an attack on the full MD5 hash function, it was close enough for cryptographers to recommend switching to a replacement, such as SHA-1 or RIPEMD-160.
- Recently, a number of projects have created MD5 rainbow tables which are easily accessible online, and can be used to reverse many MD5 hashes into strings that collide with the original input, usually for the purposes of password cracking. However, if passwords are combined with a salt before the MD5 digest is generated, rainbow tables become much less useful.

So we thought of designing a new algorithm for calculating message digest.

## II. OUR PROPOSED ALGORITHM

*Input*: A message in plaintext of any length.
*Output*: A message digest of 128 bit for the input message.
The algorithm works in the following steps:

*Step 1*

Enter the input string.

*Step 2*

Find the equivalent binary string.(Use ASCII conversion for each character used in the message string.)

*Step 3*
 i. Append a bit sequence (In this case "01") to the binary string so that the length of the resulting string is 64 shorter than a multiple of 512.
 ii. Append 64 more bits by scanning the binary string of step3(i) starting from an arbitrary location (In this case a rule can be implemented to define the starting point as [ length of the string/3 ])

*Step 4*
 i. Divide the output binary string of Step3 in 128 bit blocks.
 ii. Generate a 128 bit binary key using a random number generator.
 iii. Perform a bitwise operation ( like OR , AND, XOR, followed by Left Shift, Right Shift, zero fill shifting etc.) among the 128 bit block and 128 bit random key.
 iv. Store the output of Step4(iii) as stepwise message digest.

*Step 5*
 i. Perform a bitwise operation among the current stepwise message digest and the previous stepwise message digest.
 ii. Go to Step4 until all the blocks of input message are exhausted.

*Step 6*

Convert the output of Step5 into corresponding character value and store it as the final message digest.

*Discussion:*

The input message can be of any type( null string is also accepted).It may contain character, digit, special symbols etc.

$Message = ((a-z) + (A-Z) + (0-9) + (! - ;))^{*}$

Binary equivalent of input message then calculated. This is done by using ASCII code for corresponding character.

For example:  $A = 01000001$
 $B = 01000010$
 ……………………..etc.

The resulting binary string may contain any No. of digits. Then a bit sequence is appended (say "01") to the input binary string so that the length of the resulting string is 64 shorter than a multiple of 512.
*Mathematically    length of binary string%512=448*
The bit sequences can also be 11, 10, 00 etc.

Again 64 more bit sequence is appended to it from the input binary string itself starting from a arbitrary position. Here it starts from ([length of string]/3)$^{th}$ position.

Now the binary string is of length which is a integer multiple of 512.this message string is then divided into blocks of 128 bits.

A key is generated which is also of 128 bit. The key can be generated by using any of the random number generation method. Here we have used the method as follows
$Key = (key*39) \% 967$
$Keyf = (key)!$
Binary equivalent of keyf is calculated and any 128 bit is taken as the key for a step.

This key generation is a reoccurring function.

In each step one block of message string and a key is processed to give a message digest. The process may involve operations like OR, AND, XOR, left shift, right shift etc. The stepwise message digests performs a bitwise operation with previous step message digest to form the final message digest in binary form. The resulting binary string is then converted into character string to get the final message digest. It is to be noted that for any input string the output message digest is of constant length. The message digest may contain any characters, digits, special characters etc.

i.e.:  $Message\ Digest= ((a-z) + (A-Z) + (0-9) + (! - ;))^{n}$ ,
 *Where n= a fixed natural number*

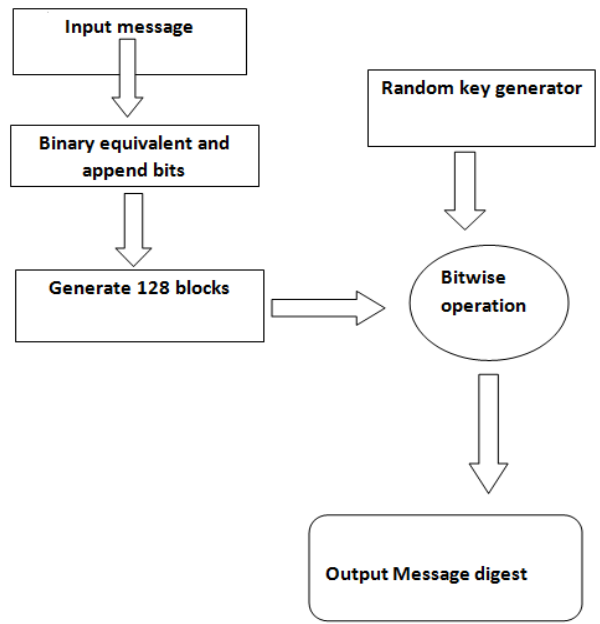

Fig. 1  A rough layout of our proposed algorithm

*Theoretical Improvements:*

- Ref. step 3.i Unlike md5 a bit sequences of "01" is appended so that the length of the resulting string is 64 shorter than a multiple of 512.the bit sequences can also be 11, 10,00 .this minimizes the No. of iterations .

- Ref. step3.ii 64 more bits are appended to the input string taking 64 bits from the input sting starting from an arbitrary position(in our algorithm [ length of the string/3]).This adds some more randomness.

- Ref. step 4.i Unlike md5 no buffer memory is used to store the 128 bit blocks. Here the whole input string is divided into 128 bit blocks without using any buffer.

- Ref. step 4.ii A random key is generated which makes it more immune to any collision vulnerability and attack.

- The output message digest contains all alphabets (lower case, upper case), digits, and special symbols.

## III. EXPERIMENTAL ANALYSIS

### A. Experimental Setup

As we know we need to enter a plaintext message to calculate the message digest so we need to design a system to take as input a plaintext message. For this purpose we have used a high level programming language, python and implemented our algorithm using this language. For drawing a comparison we have also considered other existing algorithms and their existing implementations.

We need to specify the message in plaintext as the console input. The algorithm calculates the hash digest of the message and echoes the message digest as the console output.

### B. Distinctive Analysis

Analyzing any 5 random strings and calculating the percentage of distinct characters we have drawn a table as follows:

TABLE I
DISTINCTIVENESS ANALYSIS TEST CASE

| Input String | MD5 | Our Proposed Algorithm |
|---|---|---|
| ab | 62.5 | 65.6 |
| system simulation | 50 | 59.37 |
| cd | 46.87 | 65.62 |
| project reports | 50 | 70.3 |
| niharjyoti | 40.62 | 76.56 |
| **AVERAGE** | **49.99** | **67.49** |

From the above table a comparison graph is drawn below:

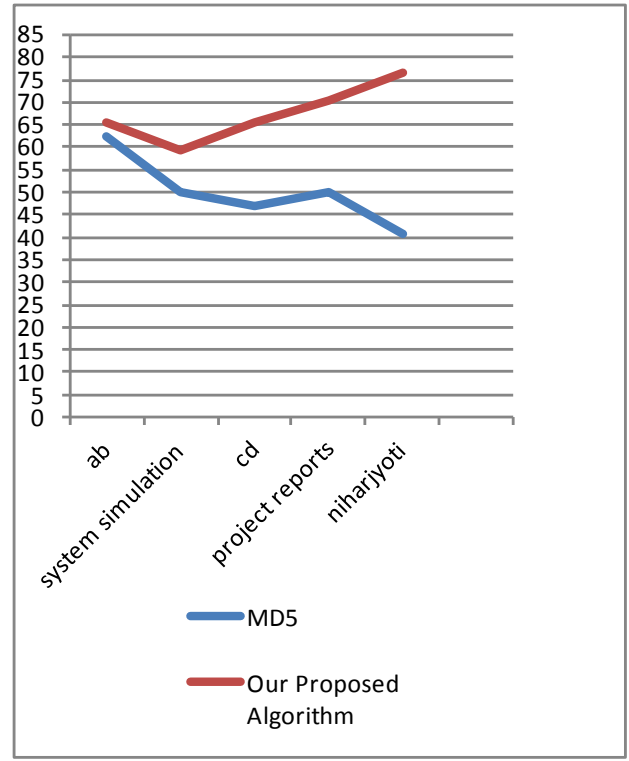

Fig. 2 A line graph which contrast the 2 algorithms on a distinctiveness basis.

### C. Probabilistic Analysis

We can measure the efficiency of our proposed algorithm and compare it with existing MD5 algorithm by considering the maximum number of characters each can have in their message digest hash value.

*A MD5 hash digest is of the general form:*
$((a-z) + (0-9))^{32}$
*So the sample space of the characters in the combination= 26+10= 36*

*Our proposed algorithm is of the general form: $((a-z)+(A-Z)+(0-9)+(!-;))^{64}$
So, the sample space of the characters in the combination= 256*

In MD5 for 32 positions the sample space of occurrence of characters is 36. So, if there is an equal distribution each position can accommodate 1.125 distinct characters.

But, in our proposed algorithm the sample space of occurrence of characters is 256. So, for equal distribution each position can accommodate 4 distinct characters.

*Thus, we can write it mathematically that our proposed algorithm is (4/1.125) =* **355.55%** *random than the existing MD5 algorithm.*

## IV. SCOPE FOR FUTURE DEVELOPMENT

- The present algorithms for calculating message digests are susceptible to brute force and rainbow table based attacks. We can further improve these classes of algorithms such that they can't be decrypted by any ordinary means practically.
- We can also work more on these algorithms so as to improve the space and time complexity further.
- The algorithm we have proposed and implemented shows a randomness efficiency of around 70 percent. We can further improve these efficiency in the future works.
- The use of MD5 in some websites' URLs means that any search engine (Google) can also sometimes function as a limited tool for reverse lookup of MD5 hashes. We can design tools like the modified version of present day salts to improve these.

## V. CONCLUSIONS

In this paper we have studied various message digest algorithms and specifically the existing MD5 algorithm. We have proposed a new algorithm, which we have shown is more random and more distinctive than the existing algorithms. We have drawn graphs, analyzed, and compared the MD5 algorithm with our algorithm by experimentation.

Our future works will be based on this project work and we will use the results obtained from this project work to design, develop and implement a more stable, robust algorithm with less time complexity.